\documentclass[12pt]{article}
\usepackage{epsfig}

\newcounter{saveeqn}
\newcommand{\alpheqn}{%
  \setcounter{saveeqn}{\value{equation}}%
  \setcounter{equation}{0}%
 \renewcommand{\theequation}{%
   \mbox{\arabic{saveeqn}\alph{equation}}}}%
\newcommand{\reseteqn}{\setcounter{equation}{\value{saveeqn}}%
  \renewcommand{\theequation}{\arabic{equation}}}
\newenvironment{mathletters}{\stepcounter{equation} %
   \alpheqn} {\reseteqn} 
\begin{document}
 \noindent
\begin{center}
  \textbf{ International Workshop on Electron Correlations and
    Material Properties, Crete Greece, June 28 - July 3, 1998}\\[4pt]
  \textit{to appear in the Conference Proceedings, Eds. N Kioussis and A.
    Gonis, Plenum Press}
\end{center}
\vspace*{60pt}
\noindent
\textbf{\large STRONGLY CORRELATED ELECTRONS: \\[4pt]
  DYNAMICAL VERTEX RENORMALIZATION}\\[20pt]
\begin{tabbing}\hspace*{1in}\=\kill
\> V\a'aclav Jani\v{s}\\[12pt]
\> Institute of Physics, Academy of Sciences  of the
Czech Republic, \\
\> Na Slovance 2, CZ-180 40 Praha, Czech Republic \\[20pt] 
\end{tabbing}
\begin{abstract}
  Single-band Hubbard model at criticality of the metal-insulator
  transition is studied using approximations derived from parquet theory.
  It is argued that only the electron-hole and interaction two-particle
  channels in the parquet algebra are relevant. A scheme is proposed how to
  reduce the parquet equations to a manageable form the complexity of which
  is comparable with single-channel approximations such as the renormalized
  RPA.  The newly derived approximation, however, contains dynamical vertex
  corrections, remains self-consistent at the two-particle level and allows
  only for integrable singularities. A qualitatively new approach for
  studying two-particle singularities at zero temperature is obtained.
\end{abstract}

\section{\normalsize INTRODUCTION}

Strongly correlated electrons and heavy fermions display a number of
phenomena not understandable within simple Fermi-liquid theory of metals.
These unusual effects include dynamical formation of magnetic local moments
and magnetic orders, mixed-valence fluctuations, Kondo and heavy-fermion
behavior, non-Fermi-liquid phases and metal-semiconductor transitions.  At
present there are two principal approaches for treating electron
correlations. Ab-initio investigation within the density-functional theory
and the local-density approximation takes explicitly into account the
Coulomb repulsion to equilibrium states within elementary cells (muffin-tin
sphere) spread over a few interatomic distances.  Short-range effects of
the electron correlations in the ground state are quite well captured. Much
less reliable are the results of the local-density approximations for
long-range effects and low-energy excitations above the ground state. This
failure can be attributed to the underestimation of the Pauli exclusion
principle and long-range fluctuations spreading over thousands of lattice
constants.  The other approach to the treatment of the electron
correlations is based on tight-binding model calculations. The microscopic
structure of the electron lattice gas is significantly simplified by taking
into account only a few atomic (Wannier) orbitals. On the other hand,
large-scale effects and low-energy excitations are well described.
Presently the two approaches are more or less complementary.
 
A realistic description of the electron correlations has to join the two
existing approaches to enable a sufficiently accurate treatment of both
short as well as long distances.  However, to succeed in a more reliable
description of the electron correlations on small and large scales, it is
mandatory to pick up and maintain the relevant aspects of each of the
construction. The former supply us with realistic static properties of the
ground-state. The latter offers qualitatively accurate estimates of
dynamical fluctuations and collective responses to low-energy excitations
of strongly correlated systems.

It appears that most of the physics of strongly correlated electrons is
covered by the archetypal Anderson impurity and Hubbard lattice models with
only a local electron interaction. The weak-coupling regime is governed by a
Hartree-Fock mean field and Fermi-liquid theories. Extended systems at low
temperatures are Pauli paramagnets with smeared out local magnetic moments.
For bipartite lattices antiferromagnetic long-range order sets in at half
filling at arbitrarily small interaction.  In the strong-coupling regime
the Hubbard model at half filling maps onto a Heisenberg antiferromagnet
with pronounced local magnetic moments and the Curie-Weiss law for the
staggered magnetic susceptibility. The spectral structure is dominated by
separated lower and upper Hubbard bands and the strongly correlated system
seems insulating even in the paramagnetic phase. The most
difficult task of the model calculations is to address the transition region
between the weak and strong coupling limits.

We present in this paper a theoretical approach based on many-body
perturbation theory with sophisticated renormalizations of one as well as
two-particle Green functions (propagators). In particular we show how the
complicated algebra of two-particle multiple scatterings can be simplified
to a manageable form. The presented single-band theory can  be
extended to the multi-band case and can hence be used for a more realistic
electron-structure calculations.

\section{\normalsize TRANSITION FROM WEAK TO STRONG COUPLING IN THE
  HUBBARD MODEL}

We have at our disposal relatively reliable techniques to describe the two
extreme limits of weak and strong couplings of the Hubbard model. We can
use perturbation expansions in each of the limits. In the former case it is
the expansion in the interaction strength while in the latter in the
kinetic energy (hopping matrix). The problem is that the two expansions do
not match each other and break down before the transition regime is
reached.  The effective Coulomb repulsion becomes comparable with the
kinetic energy in the transition region and there is no apparent small
parameter at intermediate coupling.

Dynamical fluctuations control the low-temperature physics of interacting
electrons at intermediate coupling and neither weak-coupling nor
atomic-like perturbation theories are adequate.  Weak and strong-coupling
perturbation expansions with the bare interaction can no longer reflect the
relevant physics with creation and annihilation of long-living
electron-hole pairs. The electron-hole pairs start to condense and lead to
metal-insulator and magnetic transitions generating poles in
appropriate two-particle Green functions.

The failure of the existing analytic-numerical methods to describe properly
the transition between weak and strong coupling lies in an inadequate
treatment of the two-particle criticality.  While the one-particle
functions are usually renormalized due to a self-consistency in the
self-energy, the coupling constant and two-particle vertices remain
unrenormalized.  However, the two-particle criticality demands a theory with
a dynamical renormalization of the relevant two-particle Green functions.

To succeed in a quantitative description of this complex situation, it is
necessary to reformulate perturbation theory in such a way that the bare
interaction be systematically replaced with fully renormalized two-particle
functions determined from multiple two-particle scattering processes. It
means that we have to use a perturbation theory at the level of
two-particle Green functions including dynamical vertex renormalizations.

\section{\normalsize PARQUET DIAGRAMS}

The most advanced approximation for two-particle functions are the
so-called parquet diagrams. 
Parquet diagrams were introduced to describe interaction of mesons more
effectively \cite{Sudakov56}. Since then a number of attempts have been
made to utilize the nontrivial renormalization scheme of the parquet
algebra in condensed matter. Kondo effect \cite{Abrikosov64}, x-ray edge
problem \cite{Roulet69}, formation of the local magnetic moment
\cite{Weiner70} are among the most well known applications. Inability to
solve the parquet equations effectively has impeded broader application of
the method.

Parquet diagrams represent a systematic way of a summation and
renormalization of Feynman graphs for two-particle Green functions. Instead
of concentrating on the one-particle irreducible diagrams and the Dyson
equation, the parquet approach sums two-particle diagrams contributing to
vertex functions for which Bethe-Salpeter equations are constructed. The
resulting algebra is much more complicated than that of the one-particle
approximations.  There are namely three topologically inequivalent
definitions of a two-particle irreducibility. It may be defined according
to cutting an electron-hole or electron-electron pair propagation, or
according to cutting a polarization bubble shielding the electron-electron
interaction. Each possibility defines a two-particle channel of multiple
pair scatterings characterized by a different binding of independent
variables in the vertex functions.

A general two-particle quantity will be denoted in this paper as in Fig.~1.
\begin{figure}[h]
\hspace*{80pt}
\epsfig{figure=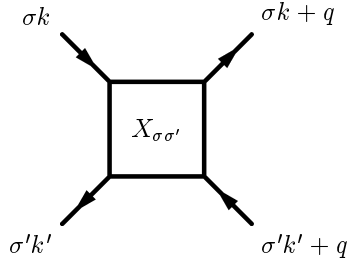,height=4cm}
\caption{\small Generic two-particle function with three independent
  four-momenta and a defined order of incoming and outgoing fermions.}
\end{figure}
Each two-particle function carries three independent four-momenta and two
spin indices. We generally denote the fermionic four-momenta $k=({\bf
  k},i\omega_n)$ and the transferred bosonic ones as $q=({\bf q},i\nu_m)$,
where $\omega_n=(2n+1)\pi T$ and $\nu_m=2m\pi T$ are the respective
Matsubara frequencies at temperature $T=\beta^{-1}$.

It is convenient to introduce a matrix notation in the spin indices to
distinguish different two-particle channels in the parquet diagrams. We
define a $2\times 2$ matrix for the generic two-particle function
$X_{\sigma\sigma'}$:
\begin{eqnarray}
  \label{eq:matrix-def}
  \widehat{X}&=&\left(\begin{array}{cc}
                 X_{\uparrow\uparrow}& X_{\uparrow\downarrow}\\
                  X_{\downarrow\uparrow}&X_{\downarrow\downarrow}
                \end{array}\right)\ . 
\end{eqnarray}
We speak about singlet and triplet contributions to a two-particle function
if the spins of the involved fermions are antiparallel or parallel,
respectively.

We define three matrix multiplication schemes for the two-particle
quantities
\begin{mathletters} 
\begin{eqnarray}
  \label{eq:conv-eh}
  \left[\widehat{X}\bullet\widehat{Y}\right]_{\sigma\sigma'}(k,k';q)&=&
  \frac 1{\beta{\cal N}}\sum_{q''} X_{\sigma\sigma'}(k,k';q'')Y_{\sigma
    \sigma'}(k+q'',k'+q'';q-q''),\\  \label{eq:conv-ee}
  \left[\widehat{X}\circ\widehat{Y}\right]_{\sigma\sigma'}(k,k';q)&=&\frac
  1{\beta{\cal N}}\sum_{q''} X_{\sigma\sigma'}(k,k'+q'';q-q'')Y_{\sigma
    \sigma'}(k+q-q'',k';q''),\\   \label{eq:conv-U}
  \left[\widehat{X}\star\widehat{Y}\right]_{\sigma\sigma'}(k,k';q)&=&\frac
  1{\beta{\cal N}}\sum_{\sigma''k''}X_{\sigma\sigma''}(k,k'';q)Y_{\sigma''
    \sigma'}(k'',k';q) 
\end{eqnarray} \end{mathletters}
representing summations over the intermediate states in the three
inequivalent two-particle channels, electron-hole, ($eh$),
electron-electron, ($ee$), and interaction, ($U$), respectively. We see
that the variables of the two-particle functions are convoluted differently
in inequivalent channels. Note that only the interaction channel mixes the
singlet and triplet contributions.

We decompose the full two-particle Green function into a sum of always
reducible and irreducible projections onto each inequivalent channel
\begin{eqnarray}
  \label{eq:K-def}
  {\cal K}_{\sigma\sigma'}(k,k';q)&=&{\cal K}^\alpha_{\sigma
    \sigma'}(k,k';q)+I^\alpha_{\sigma \sigma'}(k,k';q) 
\end{eqnarray}
where $\alpha=eh,ee,U$ refers to a two-particle channel.

The parquet diagrams can at best be represented graphically. Having in mind
the above introduced notation and a general convention that double primed
variables are summed over, we can write 
\begin{mathletters}
\begin{eqnarray}
   \label{eq:parquet-eh}&&\epsfig{figure=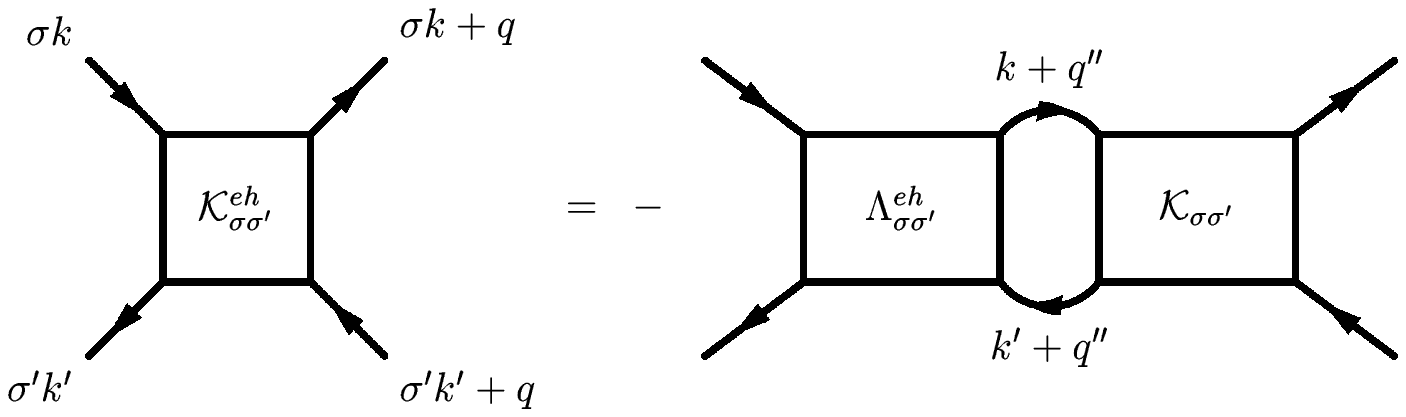,height=4cm}\\ 
   \label{eq:parquet-ee}&&\epsfig{figure=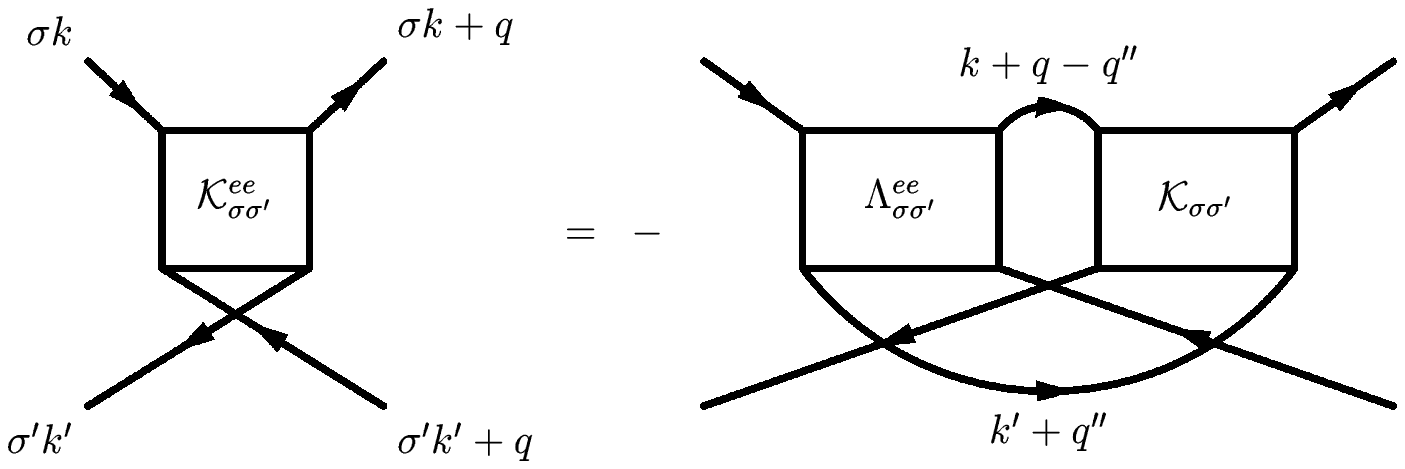,height=4.5cm}\\ 
    \label{eq:parquet-U}&&\epsfig{figure=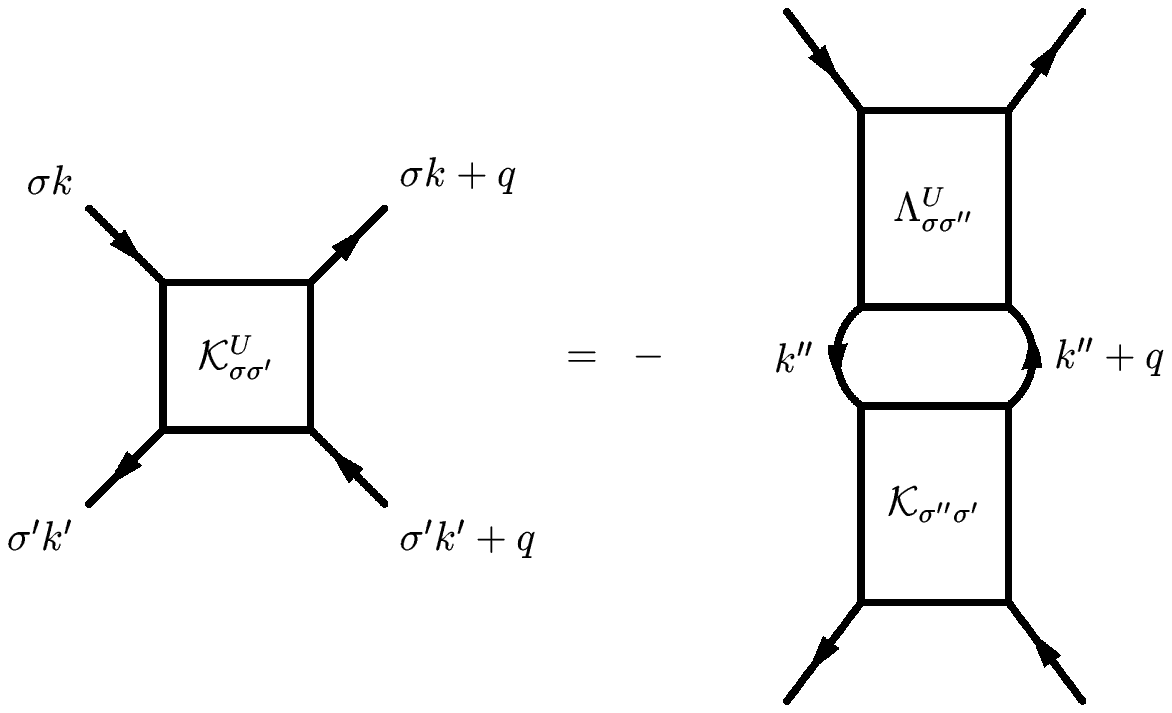,height=6cm} 
\end{eqnarray}
\end{mathletters}
Equations (4) are the Bethe-Salpeter
equations and represent one set of the full parquet approximation. To complete
it we must add relations connecting the two-particle reducible, $\cal
K^\alpha$, and irreducible, $\Lambda^\alpha$,
functions. For this purpose we introduced two-particle bubbles for each channel
\begin{mathletters}\begin{eqnarray}
  \label{eq:G2h}
 G^{(2)eh}_{\sigma\sigma'}(k,k';q)&=&G_\sigma(k+q)G_{\sigma'}(k'+q)\ ,\ 
  G^{(2)ee}_{\sigma\sigma'}(k,k';q)=G_{\sigma}(k+q)G_{\sigma'}
 (k')\ , \\[4pt] \label{eq:G2v}
  G^{(2)U}_{\sigma\sigma'}(k,k';q)&=&G_{\sigma'}(k')G_{\sigma'}
 (k'+q)\ .
\end{eqnarray}\end{mathletters}
The  functions $\Lambda^\alpha$ are then defined as
\begin{eqnarray}
  \label{eq:Lambda-def}
  \Lambda^\alpha_{\sigma\sigma'}(k,k';q)&=&I^\alpha_{\sigma\sigma'}(k,k';q)
  G^{(2)\alpha}_{\sigma\sigma'}(k,k';q)\nonumber\\
  &=&\left[U\delta_{\sigma',-\sigma}     
   +\Delta I_{\sigma\sigma'}(k,k';q) +\sum_{\alpha''\neq\alpha}{\cal K}
   ^{\alpha''}_{\sigma\sigma'}(k,k';q)\right]G^{(2)\alpha}_{\sigma\sigma'}
  (k,k';q)   
\end{eqnarray}
where $\Delta I_{\sigma\sigma'}$ is a sum of all Feynman diagrams
simultaneously irreducible in each two-particle channel. The sum contains
only higher-order diagrams where three and more particles are multiply
interconnected. In the usual treatment with only two-particle multiple
scatterings this irreducible part is neglected. We hence put $\Delta
I_{\sigma\sigma'}=0$.

Parquet equations (4) define the two-particle functions as a functional of
the renormalized one-particle propagators containing the self-energy. The
self-energy is a functional of the full two-particle function $\mathcal{K}$
from the parquet equations. Hence to complete the parquet approximation we
have to add formulas determining the dependence of the one-particle
propagator on the self-energy (Dyson equation) and the self-energy on the
full two-particle function. The consistency between one and two-particle
quantities demands
\begin{eqnarray}
  \label{eq:G-equ}
  G_\sigma({\bf k},i\omega_n)&=&\left[i\omega _n+\mu _\sigma-\epsilon
    ({\bf k}) -Un_{-\sigma } -\Sigma_\sigma ({\bf k},i\omega_n)\right]
  ^{-1}\ , \\[4pt]  \label{eq:Sigma}
  \Sigma_\sigma(k)&=&-\frac U{\beta^2{\cal N}^2}\sum_{\sigma'}\sum_{k'q}
  G_{\sigma}(k+q){\cal K}_{\sigma \sigma'}(k+q,k';-q)G_{\sigma'}(k')
  G_{\sigma'}(k'+q)   
\end{eqnarray}
where $n_\sigma$ is the particle density, $\mu_\sigma=\mu+\sigma B$, and
$\mu$ is the chemical potential and $B$ an external magnetic field.
\footnote{The self-energy $\Sigma$ in our notation measures
  corrections to the static Hartree approximation contained in the particle
  densities $n_\sigma$ determined from a sum rule $n_\sigma=
  (\beta\mathcal{N})^{-1}\sum_k G_\sigma(k)$.}

Equations (\ref{eq:K-def})-(\ref{eq:Sigma}) constitute the parquet
approximation. In this approximation only fully renormalized one and
two-particle functions appear. It means that not only all one-particle
propagators are fully renormalized with the self-energy insertions but also
the bare electron-electron interaction in the two-particle scatterings is
replaced by  corresponding irreducible two-particle (vertex)
functions.  What is missing in the parquet diagrams are three-particle
and higher-order cumulant functions. They are, however, not expected to be
relevant for the static Coulomb interaction. They contribute via dipole and
higher-order multi-pole interactions and are weak in the Hubbard model.

\section{\normalsize SIMPLIFIED PARQUET ALGEBRAS}

Parquet diagrams differ significantly from simpler, single-channel
approximations using multiple two-particle scatterings such as $RPA$, $GWA$
or $T$-matrix (TMA). The difference manifest itself in the effective interaction
in the two-particle scattering processes. The bare interaction in the
two-particle scattering events remains unrenormalized in the single-channel
approximations. In the parquet equations it is replaced by renormalized
vertex functions representing a dynamically screened interparticle
potential.  A qualitative difference between a solution to the full parquet
equations and to the single-channel approximations becomes evident at the
two-particle criticality. The vertex functions from one or more
channels get critical and sharply peaked around the Fermi energy. The
exchange between the scattered particles becomes strongly energy
dependent and cannot be approximated by a
 static effective interaction. Moreover, the
singularity at the critical point in the parquet approximation must be
integrable, while the single-channel approximations usually display
logarithmic divergences in the integrated quantities (e.~g. self-energy).

On the other hand the nonlinear integral equations represented by the
parquet diagrams are extremely complicated with an intricate analytic
structure of the solution. Each two-particle function entering the parquet
equations contains three independent (complex) variables. Neither of them
can simply be neglected, since the variables are mutually interconnected
due to the scatterings in different inequivalent channels. Consequently, a
nonperturbative solution to the parquet equations has not yet been
found. That is also why the parquet diagrams are relatively little used in
condensed matter although being in existence more than thirty years.

The only hope that the parquet approximation may become useful for 
quantitative calculations of the effects of the electron correlations is that
one succeeds in simplifying the full parquet algebra to a manageable form.
To this purpose we have to decide where we want the parquet approximation to
be reliable. As discussed in the Introduction it is the intermediate
coupling with critical two-particle functions where we need to improve upon
easy-minded approximations. We hence demand asymptotic reliability of the
parquet diagrams at the two-particle critical points and in particular at the
metal-insulator transition. We retain only the diagrams making the
effective mass of the electrons divergent at the metal-insulator
transition.

\subsection{\normalsize Two-Variable Two-Particle Functions}

It was recently argued that at least at half filling only two of the three
inequivalent channels are relevant in the critical region of the
metal-insulator transition \cite{Janis98a}. Only multiple scatterings in
the electron-hole and interaction channels contribute to divergent diagrams
at the critical point. This simplification, called dipole approximation,
however, does not reduce the number of the variables in the two-particle
functions.  It only simplifies the algebra of the parquet approximation. To
lower the number of the relevant variables we utilize the fact that the
singularities in the two-particle functions in the parquet approximation
must be integrable. It means that when integrating over the variable in
which the singularity arises we obtain a finite result. Our approximation
consists in neglecting  all finite (nondiverging) contributions at the
critical point and keeping only the divergent ones. This must be done at
the level of the diverging two-particle functions.

There is always one relevant bosonic variable in each two-particle channel.
It is the variable remaining untouched in the scattering processes. It can
easily be seen from multiplication rule (2) that it is $k-k'$, $k+k'+q$,
and $q$ for the electron-hole, electron-electron, and interaction channels,
respectively. Divergences in the reducible functions appear in the
respective conserved variables. For the metal-insulator transition only
$k-k'$ and $q$ are relevant.

In the first step we reduce the number of the relevant variables in the
two-particle functions to two. This is achieved when we neglect mixing 
of the reducible functions from different channels. We replace the full
two-particle function $\mathcal{K}$ on the right-hand side of (4) by
$U+\mathcal{K}^\alpha$, i.~e. by the sum of the completely irreducible
vertex and the reducible function in the $\alpha$-channel. Such a
replacement does not change the (universal) critical behavior of the
resultant reducible function $\mathcal{K}^\alpha$.  The bare interaction is
a constant and consequently the solution does not depend on the outgoing
variables. Further on, the triplet functions can be eliminated from this
approximation, since $I_{\sigma,\sigma}$ gets irrelevant. If we denote
$I^v_{\sigma,-\sigma}=U+\mathcal{K}^{eh}_{\sigma,-\sigma}$ and
$I^h_{\sigma,-\sigma}=U+\mathcal{K}^U_{\sigma,-\sigma}$, the parquet
equations reduce to
\begin{eqnarray}
  \label{eq:parq2a}
  I^v_{\sigma,-\sigma}(k;q)&=&U\nonumber\\ &&\hspace*{-10pt} -\frac
  1{\beta{\cal N}}\sum_{q'}I^h_{\sigma,-\sigma}(k;q')G_\sigma(k+q')
  G_{-\sigma} (k+q'+q) I^v_{\sigma,-\sigma}(k+q';q)\ , \\[4pt]
  \label{eq:parq2b} 
  I^h_{\sigma,-\sigma}(k;q)&=&U+\frac 1{\beta^2{\cal N}^2}\sum_{q',q''}
  I^v_{\sigma,-\sigma}(k;q')G_{-\sigma}(k+q') G_{-\sigma}(k+q'+q)
  \nonumber\\
 &&\hspace*{-10pt} \times I^v_{\sigma,-\sigma}(k+q';q''-q') G_\sigma
 (k+q'') G_\sigma (k+q''+q)I^h_{\sigma,-\sigma}(k+q'';q)\ . 
\end{eqnarray}
The bosonic variable $q$ has different meaning in the functions $I^h$ and
$I^v$. It is the horizontal, vertical transfer momentum for the former and
the latter, respectively. 

Because of breaking of the symmetry between the incoming and outgoing
variables in the two-particle functions we must be careful in deriving the
equation for the self-energy. To be consistent we use a symmetrized formula
with the bare interaction at the incoming and outgoing momentum. We obtain
\begin{eqnarray}
  \label{eq:Sigma2}
  \Sigma_\sigma(k)&=&-\frac U{2\beta^2{\cal N}^2}\sum_{k',q} G_\sigma(k+q)
  G_{-\sigma}(k')  G_{-\sigma}(k'+q)\left[I^h_{\sigma,-\sigma}(k+q;-q)\right.
    \nonumber\\ 
   &&\left. + I^h_{-\sigma,\sigma}(k';q)+ I^v_{\sigma,-\sigma}(k+q;k'-k)+
    I^v_{-\sigma,\sigma}(k';k-k') -2U\right]\ .
\end{eqnarray}

The reduced two-channel parquet approximation with two-variable
two-particle functions has the critical behavior of the full parquet
approximation. The universal quantities derived from the divergent functions of
the parquet equations (4) are reproduced in equations
(\ref{eq:parq2a})-(\ref{eq:Sigma2}). Only nonuniversal quantities such as
the critical interaction do not come out in this simplification 
as in the full parquet approximation.

We achieved a simplification of the full parquet algebra without loosing
leading critical behavior of the solution, but the resulting equations
have nonlinear convolutive character. We have to find a
nonperturbative solution to these equations when addressing the
metal-insulator transition. This is, however, still beyond the reach of our
analytic-numerical tools. We must find a further reduction of complexity of
the parquet equations to make them useful for a quantitative analysis.

\subsection{\normalsize One-Variable Two-Particle Functions}

It is the bosonic variable $q$ in the two-particle functions $I^h,I^v$ that
is relevant at the critical point and in which the singularity arises. The
fermionic variable $k$ in both functions labels the eigenvalues of the
integral kernel in equations (\ref{eq:parq2a}), (\ref{eq:parq2b}). The
maximal eigenvalue governs the critical behavior. If we are interested only
in the critical asymptotic behavior we can approximate the actual maximal
eigenvalue by a value at the lowest-lying fermionic four-momentum. By such an
approximation we neglect mixing of the fermionic four-momenta in the
equations and replace  $k$-dependence in the two-particle functions by a
constant, their value at $k=0$. This approach corresponds to a low-energy
expansion used by Hamann \cite{Hamann69} by assessing the strong-coupling
limit of the renormalized RPA of Suhl in the single-impurity problem.  

The suggested simplification is possible only at zero temperature, where
the difference between the fermionic and bosonic Matsubara frequencies
vanishes. We hence put $T=0$, which is the most interesting case for the
metal-insulator transition. Using the above ansatz, neglecting 
$k$-dependence in $I^h,I^v$, and putting $k=0$ in the one-electron
propagators we reduce the parquet equations to a manageable form. We
introduce new functions 
\begin{eqnarray}
  \label{eq:Gam}
  \Gamma_{\sigma}(q)&=&\frac 1{\beta{\cal N}}\sum_{q'}
  I^h_{\sigma,-\sigma}(q') G_{\sigma}(q')G_{-\sigma}(q'+q)\\  \label{eq:K} 
  \mathcal{K}_{\sigma}(q)&=& \frac 1{\beta{\cal N}}\sum_{q'}
  I^v_{-\sigma,\sigma}(q') G_{\sigma}(q')G_{\sigma}(q'+q)\ ,
\end{eqnarray}
with the aid of which we obtain for the two-particle functions
\begin{eqnarray}
  \label{eq:I^v}
  I^v_{\sigma,-\sigma}(q)&=&\frac U{1+\Gamma_{\sigma}(q)}\ ,\\[4pt]
  \label{eq:I^h} I^h_{\sigma,-\sigma}(q)&=& \frac U{1-\mathcal{K}_{-\sigma}(q)
  \mathcal{K}_{\sigma}(q)} \ .
\end{eqnarray}
The above equations have the complexity of the single-channel
approximations, but contain the full vertex renormalization in the
two-particle scatterings.  It is always the renormalized vertex
function that enters the denominator of equations (\ref{eq:I^v}), (\ref{eq:I^h})  via
the functions $\Gamma$ and $\mathcal{K}$. The single channel approximations
replace the two-particle vertex functions $I^h,I^v$ in (\ref{eq:Gam}), (\ref{eq:K}) by
the bare interaction $U$. 

It is easy to derive the corresponding equation for the self-energy. It
reads  
\begin{eqnarray}
  \label{eq:sigma1}
  \Sigma_\sigma(k)&=&-\frac U{2\beta{\cal N}}\sum_q\left[G_\sigma(k+q)
    X_{-\sigma,-\sigma}(q) \left(I^h_{-\sigma,\sigma}(q)+I^h_{\sigma,
        -\sigma}(-q)-U\right)\right.\nonumber\\
   &&\hspace*{20pt}\left. +G_{-\sigma}(k+q)X_{\sigma,-\sigma}(q)\left(I^v
       _{\sigma,-\sigma}(q)+I^v_{-\sigma,\sigma}(-q) -U\right)\right]
\end{eqnarray}
where we introduced a single bubble
\begin{eqnarray}
  \label{eq:X-def}
  X_{\sigma,\sigma'}(q)&=& \frac 1{\beta{\cal N}}\sum_{k'} G_\sigma(k')
  G_{\sigma'}(k'+q)\ . 
\end{eqnarray}
 
Equations (\ref{eq:Gam})-(\ref{eq:X-def}) replace the original parquet
approximation.\footnote{There is a different way how to reduce the
  parquet approximation to single-variable equations \cite{Abrikosov64}.
  The other reduction is based on the so-called leading logarithmic
  divergences in single-loop diagrams. This approach is inappropriate
  in the strong-coupling limit, where such divergences cannot survive
  because of the integrability of the singularities in the parquet
  equations.} They fully determine the generating two- and one-particle
functions. All thermodynamic and spectral quantities can be calculated from
them. Since we used a number of approximate steps in the derivation of the
reduced parquet algebra, any new quantity must first be formulated within
the full parquet approximation. The simplifications are then introduced via
the generating one- and two-particle functions. In this way a consistent
theory is built up, since the parquet approximation can be viewed upon as a
conserving approximation derived from a generating functional
\cite{Janis98b}.

\subsection{\normalsize Electron-Hole Symmetry}

An exact solution as well as the full parquet approximation fulfill a number
of identities and symmetries. One of the most important features of the exact
solution is the electron-hole symmetry. It means, that the physics must not
depend on whether we use the particle or the hole picture. The electron-hole
symmetry is obeyed by the complete parquet approximation. It is not clear
whether we have not lost some of the required symmetries in course of
simplifying the parquet algebra.

We lost the symmetry between the incoming and outgoing variables in
our approximate equations. Further on we also lost the electron-hole
symmetry in only one particle of the pair in the two-particle scatterings.
The former was lost during the reduction to two-variable two-particle
functions and the latter due to the suppression of the electron-electron
channel.  These losts are not so serious if we are in the critical region of
the metal-insulator transition.  The one-sided electron-hole transformation
changes the sign of the effective interaction and hence breaks the bound
electron-hole pair.

The full electron-hole symmetry in all variables must hold for any
approximate theory, otherwise we were unable to guarantee the Fermi liquid
properties at weak coupling. The electron-hole transformation may be
defined in the parameter space of the Hubbard model as
\begin{eqnarray}
  \label{eq:eh-trans}
  \mu\to\bar{\mu}=U-\mu\ ,\hspace{10pt}B\to\bar{B}=-B \ ,\hspace{10pt}
   \epsilon_\mathbf{k} 
  \to \bar{\epsilon}_\mathbf{k}=-\epsilon_{-\mathbf{k}} &&
\end{eqnarray}
where the bar denotes a hole representation.  It is easy to check by
inspection that a solution to the reduced parquet equations
(\ref{eq:Gam})-(\ref{eq:X-def}) fulfill the following symmetry relations
\begin{eqnarray}
  \label{eq:eh-sym1}
  G_\sigma(k)&=&-\bar{G}_{\sigma}(-k) \ ,\hspace{10pt} \Sigma_\sigma(k)=
  -\bar{\Sigma}_{\sigma}(-k)\ ,\hspace{10pt} n_\sigma=1-\bar{n}_\sigma \\[4pt]
 \label{eq:eh-sym2} \Gamma_{\sigma}(q)&=&\ \bar{\Gamma}_{\sigma}(-q)\ ,
 \hspace{10pt}\mathcal{K}_{\sigma}(q)=\ \bar{\cal K}_{\sigma}(-q) \ . 
\end{eqnarray}

This electron-hole symmetry together with the analyticity of the one- and
two-electron functions is used to prove Fermi-liquid properties of
solutions to (\ref{eq:Gam})-(\ref{eq:X-def}) for sufficiently weak
interaction, i.~e. before poles in the vertex functions $I^v,I^h$ appear. 

\section{\normalsize EFFECTIVE IMPURITY MODEL}

Up to know we have worked with a very general formulation of the parquet
approximation applicable in any spatial dimension. However, recent studies
indicate that most of the effects of strong electron correlations can
qualitatively be understood already within a dynamical mean-field theory or
impurity models \cite{Georges96}. It is sufficient in many situations to
reduce the effects of strong correlations to a dynamical mean-field where
the self-energy and vertex functions are local, but frequency-dependent.
Such an approach significantly reduces the numerical complexity but keeps
much of the interesting physics of strongly correlated electrons.

We apply the simplified parquet approximation to a single impurity or a
dynamical mean field model. The fluctuations in space do not contribute to
the dynamics and the four-momenta from the general formulation of the
parquet approximation collapse to frequencies.  This simplification enables
one to perform explicitly analytic continuation from the Matsubara to the
real frequencies using contour integrals. Realizing that the reduced
equations hold for $T=0$ and introducing $\zeta,z$ for the bosonic, fermionic
complex frequencies, respectively, we can write for the two-particle
functions
\begin{equation}
  \label{eq:K-cont}
  \mathcal{K}_\sigma(\zeta)=-U\int_{-\infty}^0\frac{d\omega}\pi
  \left[G_\sigma(\omega+\zeta)\mbox{Im}\frac{G_\sigma(\omega_+)}
    {1+\Gamma_{-\sigma}(\omega_+)} + \frac{G_\sigma(\omega-\zeta)}
    {1+\Gamma_{-\sigma}(\omega-\zeta)}\mbox{Im}G_\sigma(\omega_+)\right], 
\end{equation} 
\begin{eqnarray}
  \label{eq:Gam-cont}
  \Gamma_\sigma(\zeta)&=&-U\int_{-\infty}^0\frac{d\omega}\pi
  \left[G_{-\sigma}(\omega+\zeta)\mbox{Im}\frac{G_\sigma(\omega_+)}
    {1-{\cal K}_\sigma(\omega_+){\cal K}_{-\sigma}(\omega_+)}
  \right. \nonumber\\[4pt]  &&\hspace*{80pt}\left.
    +\frac{G_\sigma(\omega-\zeta)}{1-{\cal K}_\sigma(\omega-\zeta){\cal 
      K}_{-\sigma}(\omega-\zeta)} \mbox{Im}G_{-\sigma}(\omega_+)\right] 
\end{eqnarray}
and analogously for the one-particle self-energy
\begin{eqnarray}
  \label{eq:Sigma1}
  \Sigma_\sigma(z)&=&U\int_{-\infty}^0\frac{d\omega}\pi\left\{
    G_\sigma(\omega+z) \mbox{Im}\left[X_{-\sigma,-\sigma}(\omega_+)\left(
        I^h_{-\sigma,\sigma}(\omega_+) -U\right)\right]\right.\nonumber\\
  &&\hspace*{-50pt}\left. +X_{-\sigma,-\sigma} (\omega-z)\left(I^h_{-\sigma,
    \sigma}(\omega_+) -U\right)\mbox{Im}G_\sigma(\omega_+)
  +G_{-\sigma}(\omega+z)\mbox{Im}\left[X_{\sigma,-\sigma}(\omega_+)
     I^v_{\sigma,-\sigma}(\omega_+)\right] \right.\nonumber \\
   &&\left. +X_{\sigma,-\sigma}(\omega-z)I^v_{\sigma,-\sigma}(\omega-z)
    \mbox{Im}G_{-\sigma}(\omega_+)\right\} \ 
\end{eqnarray}
where $\omega_+=\omega+i0^+$. The local propagator is for the mean field  
\begin{eqnarray}
  \label{eq:G-MF}
  G_\sigma(z)&=&\int_{-\infty}^\infty d\epsilon\rho(\epsilon)\frac
  1{z+E_\sigma -\Sigma_\sigma(z)-\epsilon}
\end{eqnarray}
and for the impurity model 
\begin{eqnarray}
  \label{eq:G-Imp}
  G_\sigma(z)&=&\left[z+E_\sigma-V^2g(z+\sigma B) -\Sigma_\sigma(z)
  \right]^{-1} 
\end{eqnarray}
where $E_\sigma=\mu+\sigma B -Un_{-\sigma}$ for the mean-field and
$E_\sigma=\mu+\sigma B -\varepsilon$ for the impurity model. In the
former $\rho(\epsilon)$ is the density of states and in the latter, the
parameter $\varepsilon$ is the energy of the impurity orbital, $V$
hybridization to the conduction electrons, and $g(z)$ the one-electron
local Green function of the conduction electrons.  Note that there is no
difference at the two-particle level between the impurity and mean-field
equations.

\subsection{\normalsize Intermediate Coupling}

We use the local formulation of the simplified parquet approximation to
demonstrate that intermediate coupling is dominated by a two-particle
criticality. We confine ourselves only to half filling and the spin symmetric case
(paramagnet) where the singularity indicates a metal-insulator
transition. We further assume a bipartite lattice,
i.~e. $\rho(\epsilon)=\rho(-\epsilon)$. 

A solution to equations (\ref{eq:K-cont})-(\ref{eq:Sigma1}) has at weak
coupling the required Fermi-liquid properties. The two-particle functions
$X,\Gamma,\mathcal{K}$ are real at the Fermi energy. Iterations starting
with $U=0$  converge to a unique solution. Such a
procedure becomes numerically unstable if we increase the interaction
beyond $U_{c0}$ defined as
\begin{eqnarray}
  \label{eq:Uc0}
  U_{c0}&=-&\left[X_0(0)\right]^{-1}, \quad  X_0(0)=-\int_{-\infty}^0
  \frac{d\omega} \pi\ \mbox{Im} \left[G^{(0)}(\omega_+)^2\right] 
\end{eqnarray}
where $G^{(0)}$ is a noninteracting one-particle propagator. The value
$U_{c0}$ corresponds to the critical interaction determined from the $RPA$
pole in the electron-hole susceptibility. Explicitly,
$U_{c0}=V^2\pi^2\rho(0)$ for a mean-field ( $d=\infty$) model with a
Lorentzian DOS (infinite bandwidth) and $U_{c0}=3\pi w/4$ for a Bethe
$d=\infty$ lattice (finite bandwidth). This interaction need not
necessarily represent a critical point.  It only indicates that we cannot
start iterations from an unperturbed solution and that weak-coupling
perturbation theories get numerically unstable. We can say that the
weak-coupling regime goes over at $U_{c0}$ to an intermediate one. The
transition is, however, smooth for effective impurity models.

At intermediate coupling the functions $I^v$ and $I^h$ approach a pole at
the Fermi energy $\zeta=0$. It means that $1+\Gamma(0)$ as well as
$1-\mathcal{K}(0)^2$ approach zero. A new small (dimensionless) scale
$\Delta=1+\Gamma(0)\approx 1-\mathcal{K}(0)^2$ arises. This scale cannot be
derived from the one-electron model parameters. It is a consequence of a
two-particle criticality. In the single-impurity case it is related to the Kondo
scale. The scale vanishes at the metal-insulator transition if it is of second
order.

The vanishing scale $\Delta$ causes nonanalyticities in the two-particle
functions $\Gamma(\omega)$ and $\mathcal{K}(\omega)$, and in the
one-electron self-energy $\Sigma(\omega)$. Since the singularity in the
vertex functions $I^v(\omega), I^h(\omega)$ must be integrable, the
functions  $\Gamma,\mathcal{K},\Sigma$ remain bounded (continuous). Only
their derivatives show divergences at the Fermi energy at the critical
point. It is straightforward to show that leading-order divergences in
(\ref{eq:K-cont})-(\ref{eq:Sigma1}) are
\begin{eqnarray}
  \label{eq:K-Gam-sing}
  \mbox{Im}\ \mathcal{K}'(0_+)&\doteq& -\frac U\pi\ \frac{\left[\mbox{Im}\
    G(0_+)\right]^2}{1 +\Gamma(0)}\ , \hspace{10pt}\mbox{Im}\
    \Gamma'(0_+)\doteq -\frac U\pi\ \frac{\left[
      \mbox{Im}\ G(0_+)\right]^2}{1-\mathcal{K}(0)^2}\ ,\\[4pt]
  \label{eq:Sigma-sing} 
  \mbox{Re}\ \Sigma'(0)&\doteq&-\frac U\pi UX(0)\ \mbox{Im}\
      G(0_+)\left[\frac 1{1 
      +\Gamma(0)}+\frac 1{1-\mathcal{K}(0)^2}\right] 
\end{eqnarray}
where the prime denotes the derivative w.r.t. the frequency variable. We
utilized the symmetry of the half-filled case, $\mbox{Re}\ G(0)=0$ in the
derivation of (\ref{eq:K-Gam-sing}), (\ref{eq:Sigma-sing}) .

The Mott-Hubbard metal-insulator transition is defined by divergence of the
effective mass \cite{Brinkman70}. Since $m^*/m\propto|\Sigma'(0)|$ we have
proved that the singularity in the two-particle vertex functions $I^v,I^h$
causes divergence of the effective electron mass and hence the Mott-Hubbard
metal-insulator transition. The way the effective electron mass diverges is
essentially determined by vanishing of the Kondo scale $\Delta$.  It
means that the Kondo scale sets an interval around the Fermi energy within
which the Landau quasiparticle picture holds not only for the one-electron
but also for the two-electron functions. The narrow quasiparticle
Kondo-Suhl peak is then a consequence of sharply peaked two-particle vertex
functions. No effective static interaction can describe the Kondo
strong-coupling asymptotics and the Mott-Hubbard metal-insulator transition
with (\ref{eq:K-Gam-sing}), (\ref{eq:Sigma-sing})  reliably.

The simplified parquet approximation describes qualitatively well the
weak-coupling Fermi-liquid regime. It also correctly binds  divergence
of the effective mass with the Kondo scale arising from  a singularity in
two-particle functions. The derived approximate theory does well on the
metallic side of the metal-insulator transition. Whether the approximation
can correctly capture the physics up to the transition point must be
decided from the dependence of the Kondo scale $\Delta$ onto the
interaction strength $U$ or better on the parameter $1-U/U_c$. The correct
behavior must reproduce the exponential law
$\Delta\propto\exp\{-U\rho(0)\}$ for the single-impurity model. We need a
deeper analysis of equations (\ref{eq:K-cont})-(\ref{eq:Sigma1}) to decide
whether this really is the case. It will be done in a separate paper.

Next we have to address the strong-coupling regime for $U>U_c$.
Fermi-liquid theory no longer holds and it is unclear how to iterate
towards a stable solution. The problem is to choose the value of the
self-energy at the Fermi level. It vanishes in the Fermi-liquid, metallic
phase but diverges in the atomic, insulating solution. What value the
self-energy acquires in a self-consistent insulating solution to
(\ref{eq:K-cont})-(\ref{eq:Sigma1}) needs to be investigated in more
details. One can, however, easily check that the simplified parquet
approximation allows for a consistent $U\to\infty$ limit. After an
appropriate scaling of the variables one recovers an atomic solution at
zero temperature. Dynamical fluctuations then enter as perturbations in
$w/U$, where $w$ is an effective bandwidth of the conduction electrons.
A perturbation expansion in $w/U$ can be a good starting point for iterations
to the full solution of equations (\ref{eq:K-cont})-(\ref{eq:Sigma1}) in
the strong-coupling, insulating phase.

\section{\normalsize CONCLUSIONS}

We demonstrated in this paper that a successful description of intermediate
and strong coupling regimes in correlated electron systems demands
sophisticated approximations with a dynamical renormalization of
two-particle vertex functions. A transition regime between the weak and
strong coupling limits is governed by the formation and condensation of
long-living electron-hole pairs resulting in poles in two-particle Green
functions. These dynamical effects cannot properly be rendered by
theories with only effective static interactions.  The
well-known single-channel two-particle approximations such as RPA, TMA, or
GWA are inadequate at intermediate coupling, since the interaction in the
multiple scatterings  remains unrenormalized. Only the full parquet
approximation with all two-particle irreducible diagrams can be expected to
capture the relevant physics at the two-particle criticality of
intermediate coupling.

The parquet diagrams lead to a qualitatively different critical
behavior than that obtained from the simple-minded single-channel
approximations.  The only singularities that can appear in the parquet
approximation are integrable ones, i.~e. convolutions of  singular and
regular functions must be finite. This feature distinguishes the parquet
theory from the other existing approximations in correlated electron
systems.

The parquet algebra is too intricate to 
allow for a nonperturbative solution. Only an appropriate reduction in the
complexity of the parquet approximation can lead to a successful
application of the method. We proposed in this paper a scheme  how to
reduce the nonlinear convolutive integral parquet equations to a system of
quasi-algebraic equations not more complicated than the single-channel
approximations. The only difference is that the effective interaction in
the scattering events in one channel is dynamically renormalized by
scatterings from other channels. 

The reduction of the parquet algebra is enabled by taking into account only
leading-order divergences at the metal-insulator transition and a
low-energy expansion around the maximal eigenvalue of the parquet
equations. The resulting equations allow only for integrable singularities
and lead to sharply peaked two-particle vertex functions in the vicinity of
the critical metal-insulator transition. They correctly bind  divergence
of the effective one-electron mass with the Kondo scale stemming from a
two-particle singularity. What still remains to do is to show how well the
simplified parquet approximation reproduces the strong-coupling asymptotics
of the Kondo scale in the single-impurity problem known from an exact
solution. Only then we can be sure we have a reliable interpolation between
the weak and strong-coupling regimes of Hubbard-like models.

\section*{\normalsize ACKNOWLEDGMENT}

This work was supported in part by the Grant No. 202/98/1290 of the Grant
Agency of the Czech Republic.

\end{document}